\documentclass[11pt,twoside]{article}
\usepackage{asp2004}

\usepackage{graphics, epsfig}
\begin{document}
\title{The effects of stellar dynamics on the evolution of young dense stellar systems
}
\author{H. Belkus, J. Van Bever, D. Vanbeveren}
\affil{Astrophysical Institute, Vrije Universiteit Brussel, Pleinlaan 2, 1050 Brussels, Belgium.\\
hbelkus@vub.ac.be, jvbever@vub.ac.be, dvbevere@vub.ac.be
}

\begin{abstract}

\noindent In the present paper we report on first results of a project in Brussels where we study
the effects of stellar dynamics on the evolution of young dense stellar systems using the 3 decades
expertise in massive star evolution and our population (number and spectral) synthesis code. We
highlight an unconventionally formed object  scenario ({\it UFO-scenario}) for Wolf Rayet binaries and
study the effects of a luminous blue variable-type instability wind mass loss formalism on the
formation of intermediate mass black holes. 

\end{abstract}
\thispagestyle{plain}

\section{Introduction}

A population synthesis code calculates the temporal evolution of a population of single stars and of 
close binaries, in regions where star formation is continuous or in starbursts. Population number
synthesis (PNS) predicts the number of stars of a certain type whereas population spectral synthesis (PSS)
computes the effects on the integrated spectrum of a population of a certain class of stars. The latter is
very useful in the case of young starbursts.  

Notice that to make realistic PNS/PSS predictions of massive stars it is essential to use evolutionary
tracks calculated with the most up to date wind rates, where we distinguish those during the core hydrogen
burning (CHB) phase, the luminous blue variable (LBV) phase, the red supergiant (RSG) phase and the core
helium burning (CHeB) phase when the star is classified as a Wolf-Rayet (WR) star. A description of our PNS
and PSS code that follows the evolution of young starbursts can be found in Van Bever and Vanbeveren (2000,
2003), and references therein. 

The effects of N-body stellar dynamics may be very important in dense stellar
systems (Portegies Zwart et al., 2004 and references therein) and we therefore started recently with the
implementation of this process in our codes. In the present paper we report first results (more details
will be presented by Belkus et al., 2005). In section 2 we describe our model. In section 3 we further
discuss the unconventionally formed object scenario ({\it UFO-scenario}) introduced by Dany Vanbeveren, these
proceedings and section 4 illustrates the effect of an LBV-type instability in the most massive stars on the
formation of intermediate mass black holes (IMBHs).

\section{The model}

Since we are interested in the spectral evolution of young stellar systems, we decided in favor of 
direct N-body integration. The integrator is written in Brussels and linked to our PSS and/or PNS code.
Since we are mainly interested in the evolution during the first few million years, to save computer time
we approximate the evolution of our cluster by generating a (large) number of massive (initial mass
between 10 and 120 M$_{\odot}$) objects from a Salpeter initial mass function and a corresponding number of
objects with a mass $\le$ 10 M$_{\odot}$. The latter objects are fixed in the cluster in space and time but
the effect on the trajectories of the massive objects is included in the N-body integration. Each massive
object of mass M can be a single star with mass M or a binary with total mass M. In case of a binary, the
mass ratio is drawn from a flat mass ratio distribution and the period is drawn from a distribution which is
constant in the Log (only binary periods smaller than 10 years so that our PNS/PSS code can handle their
evolution).  The interaction of two objects is treated with the chain regularision method as explained by
Mikkola and Aarseth (1993, and references therein). When at least one of the objects is a binary, the effect
of the direction of the impact with respect to the orbital plane and orbital phase is calculated using a
Monte-Carlo method. 

\section{A {\it UFO-scenario} for WR+OB binaries}

The formation of WR+OB binaries in young dense stellar systems may be quite different from the conventional 
binary evolutionary scenario as it was proposed by Van den Heuvel and Heise (1972). Mass segregation in
dense clusters happens on a timescale of one or a few million years which is comparable to the
evolutionary timescale of a massive star. Within the lifetime of a massive star, close encounters may
therefore happen very frequently. When we observe a WR+OB binary in a dense cluster of stars, its
progenitor evolution may be very hard to predict. Our simulations predict the following unconventionally
formed object scenario (a UFO-scenario as introduced by D. Vanbeveren in the present proceedings) of
WR+OB binaries. After 4 million years the first WR stars are formed, either single or binary. Due to mass
segregation, this happens most likely when the star is in the starburst core. Dynamical interaction with
another massive object becomes probable, especially when the other object is a binary. We encountered a
situation where an object which started as a 50 M$_{\odot}$ single star and evolved in 4 Myrs into a
single WC-type star with a mass = 10 M$_{\odot}$, interacts with an object with total mass = 30 M$_{\odot}$,
in our case a 16  M$_{\odot}$ + 14 M$_{\odot}$ circularized binary with a period P = 6 days. Since we use a
Monte-Carlo method in order to calculate the remnant after the interaction, it may be interesting to
investigate all possible remnants and their occurence frequency. The program FEWBODY (Fregeau et al., 2004) 
is very suited for this purpose. Figure 1 shows the results of 22000 simulations (with different impact
parameters). The following objects are possible. (1 and 2): The WC star merges with one of the binary
components; the merger forms a binary with the remaining star. (3): the
original binary components merge (and forms a 30 Mo rejuvenated and nitrogen enhanced star); the latter
forms a binary with the WC star. (4 and 5): exchange binaries where the WC star replaces one of the
original binary components. (6): the preservation of the situation before the interaction, but of course
different trajectories and different orbital parameters of the binary. (7): the 3 stars merge and form 1
single star. (8 and 9): one of binary components merges with the WC star; the merger and the remaining binary
component become disrupted and further evolve as single stars.    

Within the class (3), the following final object is a possibility : the two binary components merge and the
30 M$_{\odot}$ merger forms a  binary with the WC star with a period of $\approx$ 80 days and an
eccentricity e = 0.3. This binary resembles very well the WR+OB binary $\gamma$$^2$-Velorum but it is clear
that conventional binary evolution has not played any role in its formation.

In all our simulations, the resulting binaries are eccentric and we therefore conclude that when an eccentric
WR+OB binary is observed in a cluster where dynamical interaction may have occured, predicting its progenitor
evolution becomes ambiguous. 

\begin{figure}[h]
\center
\epsfig{file=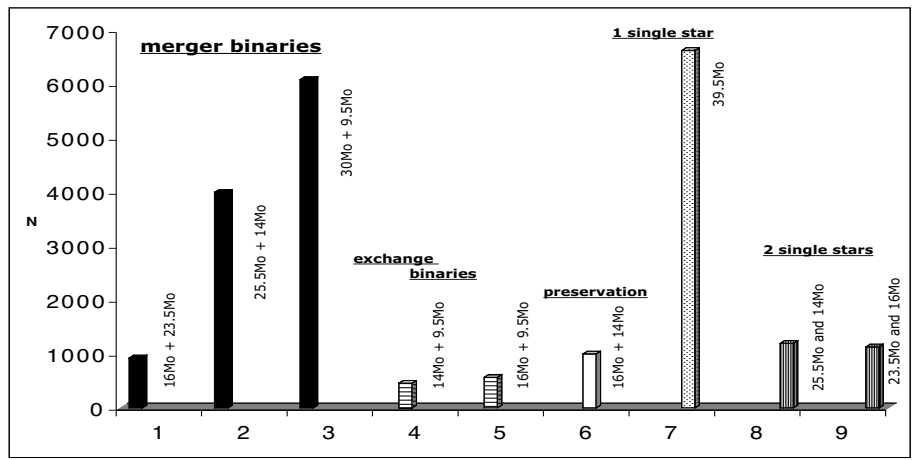, height=6cm,width=12cm}
\caption{Possible remnants and their occurence frequency within 22000 simulations after an interaction of a
10 M$_{\odot}$ WC star with a 16 M$_{\odot}$ + 14 M$_{\odot}$ circularized binary with a period P = 6 days.}
\label{fig:fig1}
\end{figure}

\section{The formation of an IMBH }

The formation of massive black holes through collision runaway in dense young star clusters has been 
studied by Portegies et al. (2004). They applied their results to the cluster MGG-11 in the starburst
galaxy M82. To follow the evolution of the massive stars, they use a scenario as explained in Portegies
Zwart et al. (1999). However, the stellar wind mass loss rate formalisms deserve some attention,
especially the formalism during the LBV phase of a very massive star as discussed by D. Vanbeveren in the
present proceedings. In order to investigate the effect on the formation of IMBHs, we generated a cluster
with 3000 massive single objects, a King (1966) distribution with parameters so that the simulation may be
appropriate for MGG-11. We first made a simulation using evolutionary calculations where the LBV-type mass
loss rate is switched off. In a second simulation, all stars (collision products) with a mass $>$ 120
M$_{\odot}$ evolve with a stellar wind mass loss rate = 10$^{-3}$ M$_{\odot}$/yr. When the mass drops below
120 M$_{\odot}$, we switch back to our normal stellar evolution as it is implemented in our PNS/PSS code (see
introduction). Collision products are mixed instantaneously and since we follow the pre-collision stars in
detail, we calculate the resulting chemical abundances of the mixed star from first principles. The further
evolution of this merger is calculated with our stellar evolutionary code with the appropriate abundances. 
Figure 3 shows two simulations, one without and one with the LBV-mass loss. In both cases a runaway
collision starts after a few 10$^{5}$ yrs. However, the simulation with the LBV-type mass loss formalism
switched on illustrates that after the runaway process there may be time enough for the merger to lose
sufficient mass so that it becomes a normal 120 M$_{\odot}$ star. 

The main conclusion of our calculations is that in order to study the possibility to form IMBHs in young 
dense stellar systems, a good knowledge of the LBV-type instability in very massive stars and the
resulting mass loss rate is essential.

\begin{figure}[h]
\center
\epsfig{file=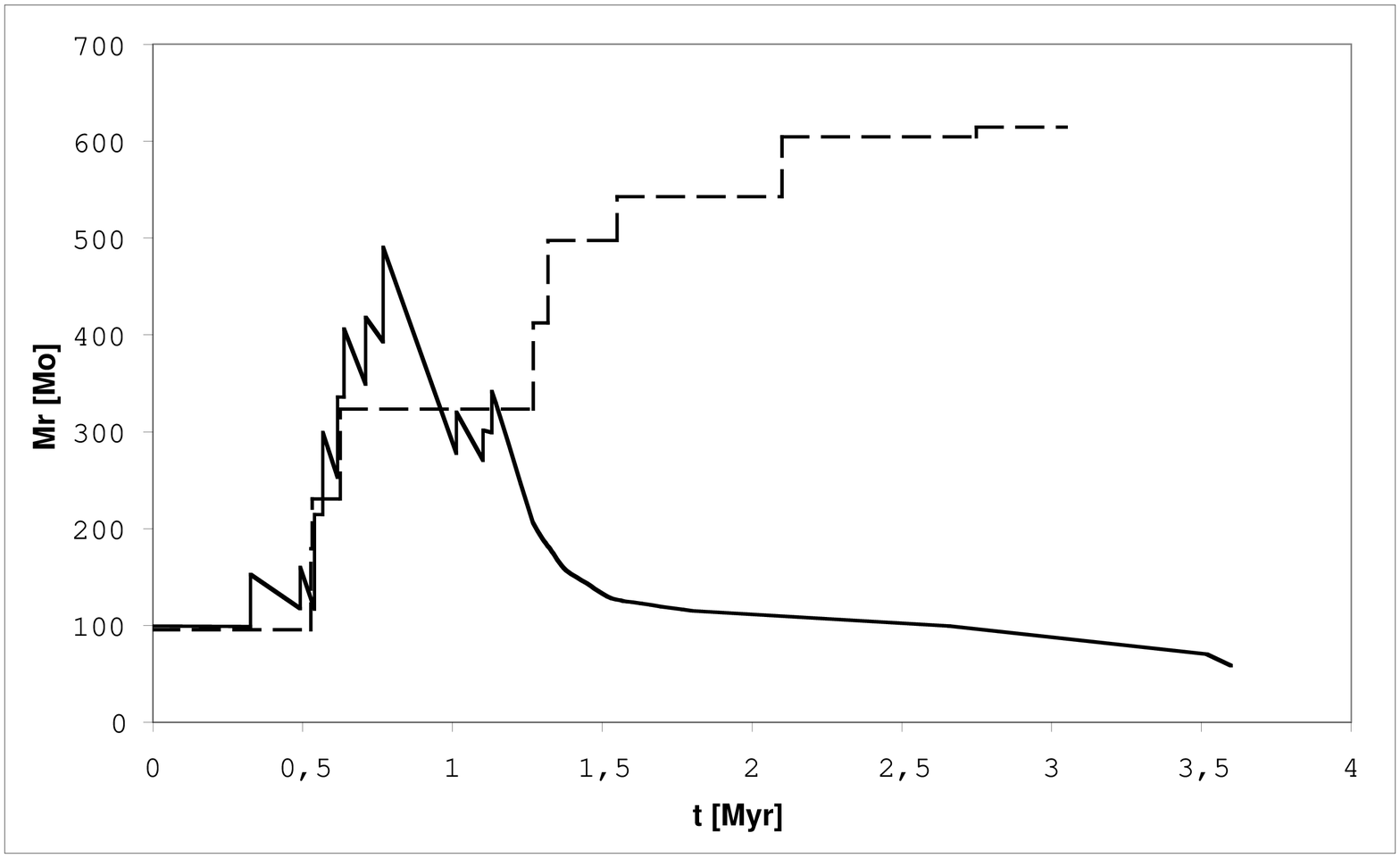, height=6cm,width=12cm}
\caption{The variation of the mass of the most massive star in the cluster. The dashed line shows the
runaway mass growth when the LBV mass loss is switched off whereas the full line illustrates the variation
of the mass of the same object when the LBV mass loss is included as explained in the text. }
\label{fig:fig2}
\end{figure}

\end{document}